\documentclass[aps,twocolumn,groupedaddress,showpacs]{revtex4}
\usepackage{graphicx}
\usepackage{dcolumn}
\usepackage{bm}
\usepackage{epstopdf}
\usepackage[dvipsnames,usenames]{color}
\usepackage{color}
\usepackage{float}
\usepackage{amsmath}
\usepackage{amssymb}
\usepackage{amsthm}
\usepackage{mathrsfs}

\emergencystretch=\maxdimen
\begin{document}
\title{High-Q plasmonic crystal laser for ultra-sensitive biomolecule detection}

\author{Jiacheng~Sun$^{1,3}$, Tao~Wang$^{1,2,\footnote{wangtao@hdu.edu.cn}}$, Zeinab Jafari$^{4}$, Fei~Gao$^{5}$, Xiao~Lin$^{5}$, Hongsheng~Chen$^{5}$, Gaofeng~Wang$^{1,2}$, Israel De Leon$^{4,\footnote{ideleon@tec.mx}}$}

\affiliation{$^1$Engineering Research Center of Smart Microsensors and Microsystems of MOE, Hangzhou Dianzi University, Hangzhou, 310018, China}
\affiliation{$^2$School of Electronics and Information, Hangzhou Dianzi University, Hangzhou, 310018, China}
\affiliation{$^3$School of Zhuoyue Honors, Hangzhou Dianzi University, Hangzhou, 310018, China}
\affiliation{$^4$School of Engineering and Sciences, Tecnol\'ogico de Monterrey, Monterrey, Nuevo Le\'on 64849, Mexico}
\affiliation{$^5$Interdisciplinary Center for Quantum Information, College of Information Science and Electronic Engineering, Zhejiang University, Hangzhou 310027, China}


\date{\today}

\begin{abstract}
Plasmonic lasers provide a paradigm-changing approach for the generation of coherent light at the nanoscale. In addition to the usual properties of coherent radiation, the emission of plasmonic lasers can feature high sensitivity to the surrounding environment, which makes this technology attractive for developing high-performance and highly-integrated sensing devices. Here, we investigate a plasmonic laser architecture based on a high-$Q$ plasmonic crystal consisting of a periodic arrangement of nanoholes on a thin gold film cladded with an organic-dye-doped SiO$_2$ gain layer as the gain material. We report an extensive full-wave numerical analysis of the device's lasing performance and its application as a biochemical sensor, showing that the proposed design features excellent figures of merit for surface sensing that in principle can be over an order of magnitude larger than those of previously reported high-performance plasmonic biosensor architectures.
\end{abstract}

\maketitle

\section{Introduction}
Surface plasmons are optical surface waves formed by the coupling of an optical field to free electrons at the surface of a metal~\cite{Maier2007}. There exists two types of surface plasmon waves, i.e., surface plasmon polaritons (SPPs) and localized surface plasmons (LSPs). The former type refers to surface plasmon waves that propagate along open metallic surfaces, while the latter refers to standing waves confined to nanoscopic metallic structures. Their tight field confinement to the metal's surface~\cite{Barnes2003, Schuller2010} enables these surface waves to probe minuscule fluctuations of the refractive index in the vicinity of the metal surface~\cite{Lal2007, Anker2008, Brolo2012}. As such, surface plasmons have been extensively exploited for sensing applications. In particular, plasmonic biosensor technology has been widely used for biochemical applications because it offers high sensitivity, as well as capacity of real-time, label-free quantitative analysis on a broad variety of analytes, ranging from chemicals, through proteins and nucleic acids, to bacterial and viral pathogens~\cite{Lal2007, Brolo2012, Homola2006, Spackova2016, Dahlin2013, Jackman2017}. Numerous plasmonic biosensor architectures have been proposed and implemented throughout the years. Perhaps the most popular one is the surface plasmon resonance (SPR) architecture, a mature technology that makes use of the SPP supported by a thin metallic film~\cite{Homola2006}. Other more sophisticated architectures with chip-integration capacity include the use of SPPs in various waveguide configurations~\cite{Dostalek2001, Krupin2013, Wong2015, Gao2011}, and LSPs supported by metallic nanoparticles or nanoholes in metallic films~\cite{Anker2008, Spackova2016}. In addition, grating-like structures consisting of periodic arrays of plasmonic nanoparticles or nanoholes, typically known as plasmonic lattices or plasmonic crystals have recently attracted considerable attention as sensing platforms because they can support delocalized resonances with significantly higher quality factors ($Q$ factors) and promising sensing figures of merit (FOM)~\cite{Garcia2007, Garcia2010, Kravets2018, Vala2019}. Indeed, the advances in plasmonic biosensor technology thus far represent a remarkable progress. Yet, improving the sensitivity and FOM of plasmonic biosensors is still of great importance as it would expand their scope of applications; for instance, achieving real-time, labelfree single molecule detection, highly-integrated point of need systems, and robust clinical diagnosis ~\cite{Zijlstra2012, Chen2011, Liu2020, Mauriz2019}.

Referring specifically to the sensor's transducer, the performance of all previously mentioned sensing architectures is ultimately limited by the optical losses of the system, which reduces the resonance's $Q$ factor, thus limiting the interaction with the analyte~\cite{Dastmalchi2016} and degrading the device's FOM. Hence, a compelling approach to improve the sensing performance is to reduce or completely mitigate the plasmonic losses by incorporating optical gain in the system. Indeed, surface plasmon amplifiers and lasers have been demonstrated in the laboratory~\cite{Berini2012, Wang2017, Wu2019} and their prospects for applications are vast~\cite{Ma2019}. In particular, recent theoretical and experimental investigations have indicated that active plasmonic biosensing architectures can offer a significant improvement on the sensing performance because of the narrow spectral linewidth achieved in these systems~\cite{Ma2014, Wang2017a, Zhu2017, Cheng2018}. While these pioneering investigations have reported an improved FOM over previously reported passive devices, more efforts are still required to demonstrate an active biosensor that satisfy key factors required for ultra-sensitive biomolecule detection, in particular the possibility of achieving a large FOM for bulk and surface sensing using a platform capable of operating at room temperature.

In this work, we investigate the sensing performance of an active plasmonic biosensor configuration based on a high-$Q$ plasmonic crystal incorporating an organic gain medium. The structure consists of a thin gold film decorated with a two-dimensional periodic array of nanoholes and cladded with an optically-pumped Rhodamine 640-doped glass layer. The high-$Q$ plasmonic Bloch mode supported by the passive structure favours plasmonic lasing with gain levels previously demonstrated in organic media at room temperature~\cite{Suh2012, Yang2015}. We characterize the optical properties and sensing performance of the proposed structure through extensive finite-difference time domain (FDTD) simulations incorporating the rate equations governing the time domain dynamics of the gain medium's atomic populations. Our analysis indicates that lasing with a narrow linewidth of 0.24 nm is possible under practical conditions. Using wavelength interrogation, we show that such a narrow linewidth together with a large sensitivity result in FOM for bulk sensing as large as $\sim$ 1000 RIU$^{-1}$ and a FOM for surface sensing of $\sim$ 65 RIU$^{-1}$ for thin biolayers in the order of 5 nm. The bulk FOM, in this work, is superior to those experimentally reported in previous passive and active works based on wavelength interrogation~\cite{Spackova2016, Wang2017a, Wang2017b}, and comparable to values predicted recently for plasmonic laser-based systems~\cite{Zhu2017, Cheng2018}, The surface FOM is also at least one order of magnitude larger than that of passive plasmonic biosensors~\cite{Spackova2013, Spackova2019}, indicating that this sensing architecture has potential for ultra-sensitive biomolecule detection.

\section{Principle of operation and structure analysis}
The optical properties of passive plasmonic crystals consisting of thin metallic films patterned with a two dimensional array of holes have been studied extensively in the past both from the fundamental point of view~\cite{Garcia2010, Dickson2015} and for biosensor applications~\cite{Vala2019, Malyarchuk2005, Leebeeck2007}. In this section, we describe both the passive and active optical properties of the proposed plasmonic crystal structure.

\subsection{Passive structure}
\begin{figure}
\centering
\includegraphics[width=7.5cm,height=14cm]{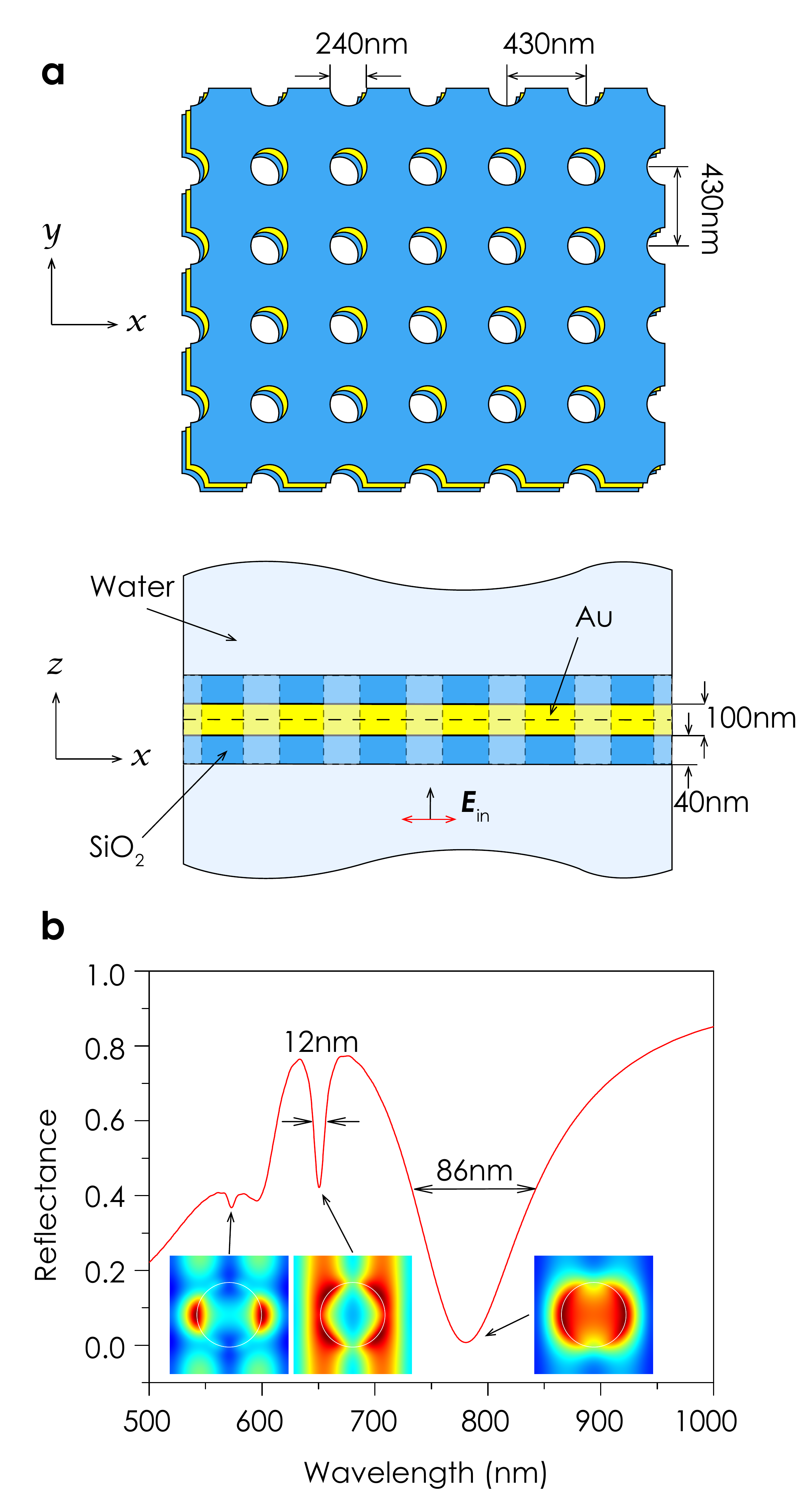}
\caption{Optical response of the passive structure: (a) Schematic illustration of the passive plasmonic crystal; the top panel shows a top view [($x$, $y$)-plane] and the bottom panel shows a side view [($x$, $z$)-plane]. (b) Reflection spectrum for
visible and near infra-red wavelength. The broad dip corresponds to the LSP resonance of the individual nanohole, while the other two narrow dips (at 648 nm and 575 nm) are plasmonic Bloch modes of the structure. The insets illustrate the field intensity over the crystal's unit cell associated with the indicated spectral dips.}
\label{structure}
\end{figure}

We consider first the passive structure, which is illustrated in Fig.~\ref{structure}a. It consists of a gold film cladded on both sides by a thin silicon dioxide (SiO$_2$) layer patterned with a two-dimensional array of circular holes arranged as a squared lattice surrounding by water. The thicknesses of the gold and each of the SiO$_2$ layers are 100 nm and 40 nm, respectively, the diameter of the holes is 240 nm, and the lattice period is a = 430 nm. Furthermore, since biochemical analytes are usually in an aqueous solution, we consider water as the medium surrounding the structure. We assume that the structure is illuminated by an $x$-polarized plane wave propagating along the positive $z$ direction that impinges at normal incidence onto the structure (see bottom panel of Fig.~\ref{structure}a).

The optical properties of the structure are determined to a large extent by the plasmonic Bloch modes supported by the film~\cite{Dickson2015}. These modes are $p$-polarized and can be categorized into \textit{symmetric} or \textit{antisymmetric} modes depending on whether the main magnetic field component has even or odd symmetry with respect to the $z$ = 0 plane, respectively. The in-plane wave vector of these Bloch modes can be approximated by \textbf{k}$_{l,m}$ = \textbf{k}$_{SP}$ + \textbf{G}, where \textbf{k}$_{SP}$ is the wave vector of the SPP supported by the structure in the absence of the holes and \textbf{G} = $l$\textbf{g}$_x$ + $m$\textbf{g}$_y$ is the reciprocal lattice vector of the crystal, with g$_x$ = g$_y$ = 2$\pi/a$ being the primitive reciprocal vectors associated with the $x$ and $y$ directions, and $l$ and $m$ are integers. For normal incidence illumination, only those modes with \textbf{k}$_{l, m}$ = 0 can be excited due to the in-plane momentum conservation. This results in the excitation of standing-wave resonances formed by Bloch modes with the same wave vector magnitude but opposite directions - i.e., modes associated with the indices [$\pm l$, $\pm m$].

With this in mind, one can identify the signature of these Bloch modes in the reflectance spectrum shown in Fig.~\ref{structure}b, obtained through FDTD calculations. The broad resonance, centered at the wavelength $\lambda = 778 nm$, corresponds to the symmetric [$l$, $m$] = [$\pm$1, 0] mode, while the other two narrow features centered at 648 nm and 575 nm correspond to the antisymmetric [$\pm$1, 0] and [$\pm$1, $\pm$1] modes, respectively. We note that the antisymmetric [$\pm$1, 0] mode features a narrow linewidth of only $\sim$ 12 nm (corresponding to a Q factor of $\sim$ 54), which results from its weak coupling to radiative modes. The symmetric mode has a much broader linewidth of $\sim$ 86 nm. Both, symmetric and antisymmetric modes have been studied for passive sensing applications, showing similar bulk sensitivities~\cite{Vala2019}. In our case we focus on the antisymmetric [$\pm$1, 0] mode because its small radiative loss facilitates lasing using an optically pumped organic dye, which is attractive for its low cost and ease of fabrication.

\subsection{Active structure}
\begin{figure*}
\centering
\includegraphics[width=16cm,height=11cm]{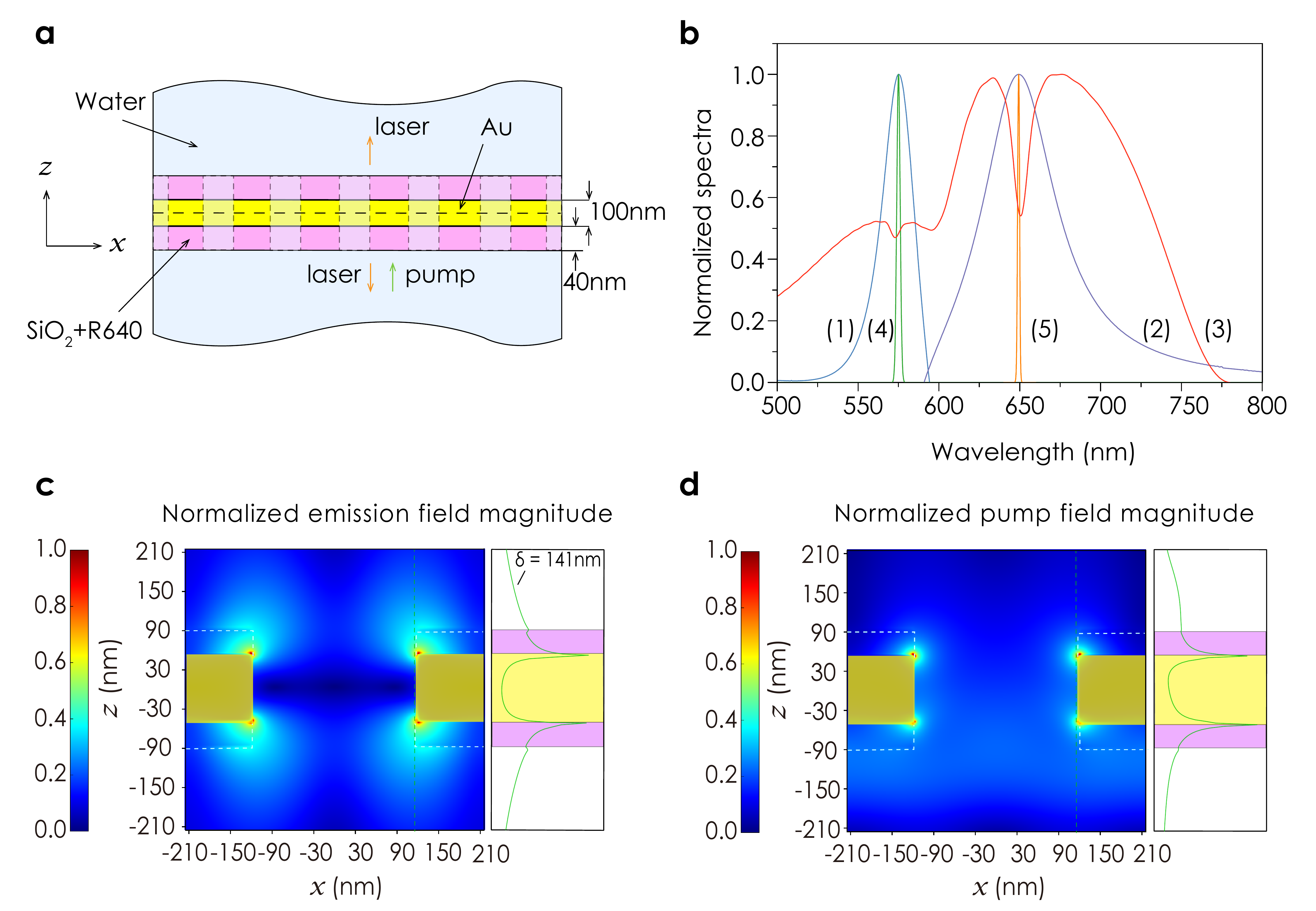}
\caption{Optical response of the active structure: (a) Schematic illustration of a cut through the ($x$, $z$)-plane of the active plasmonic crystal. All the parameters are the same as for the passive structure, except that the cladding is substituted by SiO$_2$ doped with R640 dye. (b) Relevant spectra for the active case: absorption (1) and emission (2) spectra of the R640
dye; reflection spectrum of the passive structure (3); and pump (4) and stimulated emission (5) spectra. (c) and (d) show the
normalized electric field magnitude in the ($x$, $z$)-plane through the center of the hole for the stimulated emission and pump
fields, respectively. The dashed boxes superimposed indicate the location of the gain media. The curves on the right of (c,d)
represent the field distribution evaluated at the points given by the dashed lines. The field in water in (c) decays exponentially with a decay constant $\delta$ = 145 nm.}
\label{Active}
\end{figure*}

A side view of the active structure is shown in Fig. Fig.\ref{Active}a. In this case, we have replaced the SiO$_2$ claddings with SiO$_2$ doped with Rhodamine 640 (R640) molecules, while keeping all the other structural parameters identical as those for the passive case. 
To model the active structure, we used an FDTD solver incorporating a semi-classical treatment for the gain medium. The semi-classical model is based on the solution the rate equations governing the time-domain dynamics of the molecular populations in the organic gain medium~\cite{Chang2004}. The gain medium is treated as a four-energy-level system with Lorentzian emission and absorption spectra peaking at 640 nm and 570 nm, respectively. For all our simulations we have used a molecular concentration of N = 4.5 $\times$ 10$^{19}$cm$^{-3}$, and assumed that the dipole moment of such molecules is randomly oriented in space. The molecular parameters of R640 relevant for our calculations, including the radiative lifetime (4 ns) and emission spectral linewidth (300 THz), were selected according to values previously reported in the literature for a similar molecular concentration in SiO2~\cite{Dardiry2011}. Detailed information about the FDTD model and R640 parameters can be found in the supplementary material. 

The absorption (1) and emission (2) spectra of R640 used in the FDTD simulations, as well as the reflectance spectrum of the passive plasmonic crystal (3) are illustrated in Fig.~\ref{Active}b. The passive structure was optimized to support the antisymmetric [$\pm$1, 0] mode and the [$\pm$1, $\pm$1] modes close to the peak of the emission and absorption spectra of R640, respectively, in order to facilitate the interaction of these two modes with gain medium. We have used a 200 fs pump pulse with center wavelength $\lambda_p$ = 575 nm impinging at normal incidence (see Fig.~\ref{Active}a), thus exciting the [$\pm$1, $\pm$1] Bloch mode of the structure. As shown in Fig.~\ref{Active}c and Fig.~\ref{Active}d, this approach allows us to concentrate the pump field tightly to both surfaces of the metal film, and obtain good overlap between the pump field (the [$\pm$1, $\pm$1] mode), the emission field (the [$\pm$1, 0] mode), and the gain medium, which is crucial to achieve lasing.

\begin{figure}
\centering
\includegraphics[width=7cm,height=12cm]{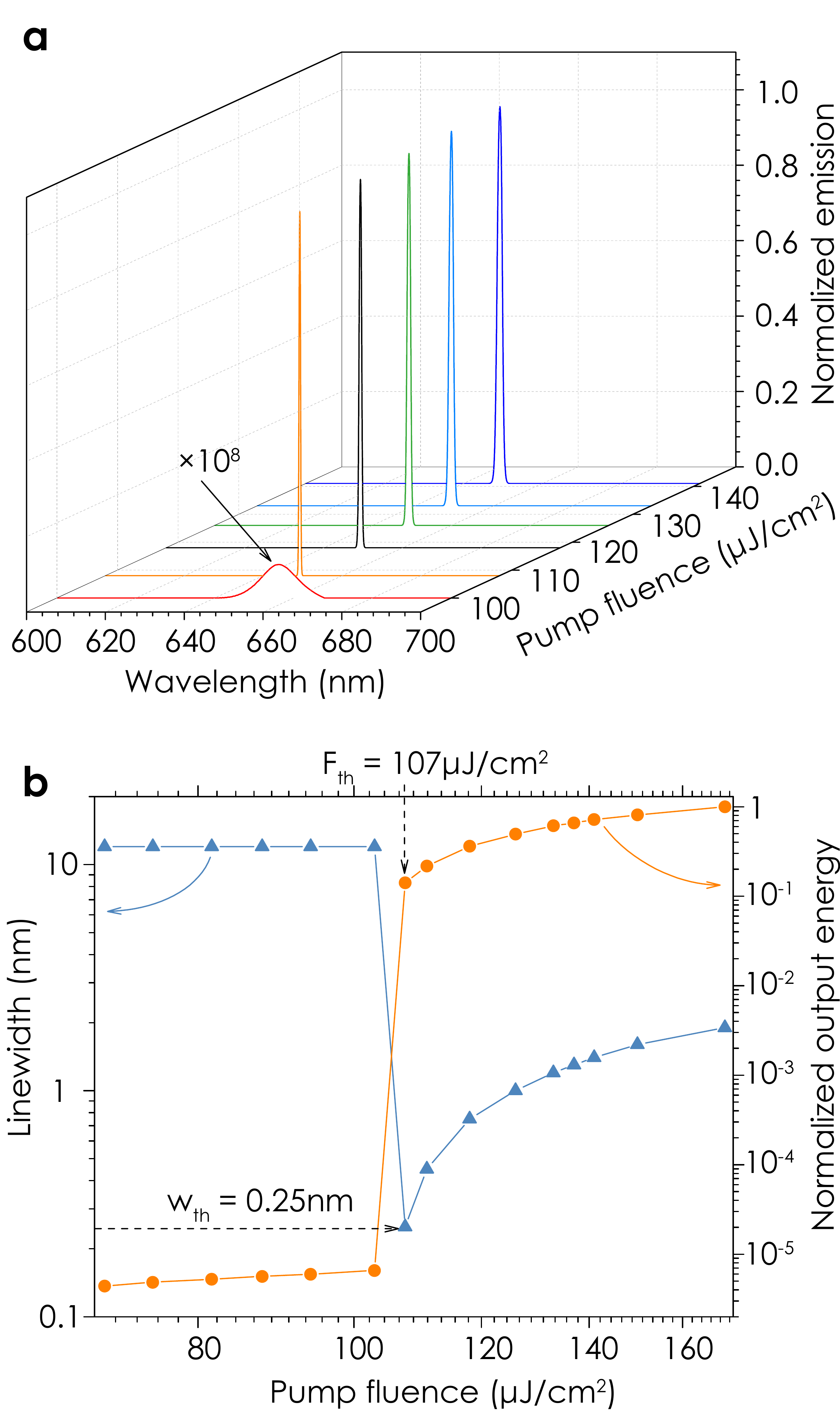}
\caption{Emission characteristics of the active structure: (a) Evolution of the emission spectrum as a function of the pump
fluence. (b) Normalized peak value (blue circles) and spectral linewidth (orange triangles) of the emission spectrum as
the function of pump fluence. The pump threshold for laser emission is estimated as F$_{th}$ = 107 $\mu$J/cm$^2$.}
\label{Lasing}
\end{figure}

The emission spectrum of the antisymmetric [$\pm$1, 0] mode in the presence of the gain medium is calculated for a pump fluenc, $F$, varying from 70 to 170 $\mu$J/cm$^2$. The results, given in Fig.~\ref{Lasing}a, show a broad emission spectrum characteristic of spontaneous emission for low pump fluences, and a sudden increase in the emission intensity and reduction of the spectral linewidth at $F_{th}$ = 107 $\mu$J/cm$^2$, indicating the threshold for lasing. The wavelength of the laser emission at the pump threshold occurs at $\lambda_e$ = 649.35 nm, which is slightly shifted from the central wavelength of the antisymmetric [$\pm$1, 0] resonance [see spectrum (1) in Fig.~\ref{Active}b]. The linewidth and normalized peak of the emission spectrum is illustrated in Fig.~\ref{Lasing}b as a function of $F$. Remarkably, the linewidth at the threshold condition is only $w_{th} = 0.24$ nm and it gradually increases for larger $F$ values. This characteristic behavior of the spectral linewidth has been reported in several experimental investigations on plasmonic nanolasers~\cite{Ma2014, Yang2015, Hakala2017, Rekola2018}. Furthermore, similar linewidths have recently been reported experimentally for various configurations of plasmonic lasers~\cite{Zhu2017, Cheng2018, Hakala2017}.

The energy loss per unit length of the passive cavity can be obtained as $\alpha = \omega n_{g}/cQ$, where $\omega$, $c$, and $n_{g}$ are the angular frequency, speed of light in vacuum, and group index of the Bloch mode, respectively. Using the values $Q$ = 54 and n$_{g}$ = 1.19 obtained for the antisymmetric [$\pm$1, 0] mode, we obtain $\alpha = 2133 cm^{-1}$. The
the cavity. It is important to point out that experimental investigations related to both, dielectric microlasers~\cite{Campillo1991, Djiango2008} and plasmonic lasers~\cite{Suh2012, Yang2015}, have reported an enhancement in the stimulated emission rate resulting from the Purcell effect, whereby the gain provided by the molecular medium is enhanced by a factor equal to the Purcell factor of the cavity~\cite{Campillo1991}. As this effect is not completely taken into account in our simulations, and our structure exhibits an averaged Purcell factor of $\sim$ 2.2 over the gain medium (see supplementary material), we expect that the molecular concentration and pump requirements predicted by our simulations are overestimated.

\section{Sensing performance}
The performance of optical biosensors are typically evaluated based on the bulk sensitivity, (S$_B$) and surface sensitivity (S$_S$) parameters, and their respective FOMs~\cite{Spackova2016}. In particular, the S$_B$ parameter is useful to compare the performance of different biosensor configurations, while the S$_S$ parameter gives a more precise estimate of plasmonic affinity biosensors~\cite{Li2015}. In what follows, we discuss the performance of the proposed active structure for both, bulk and surface sensing.

\begin{figure*}
\centering
\includegraphics[width=16cm,height=11cm]{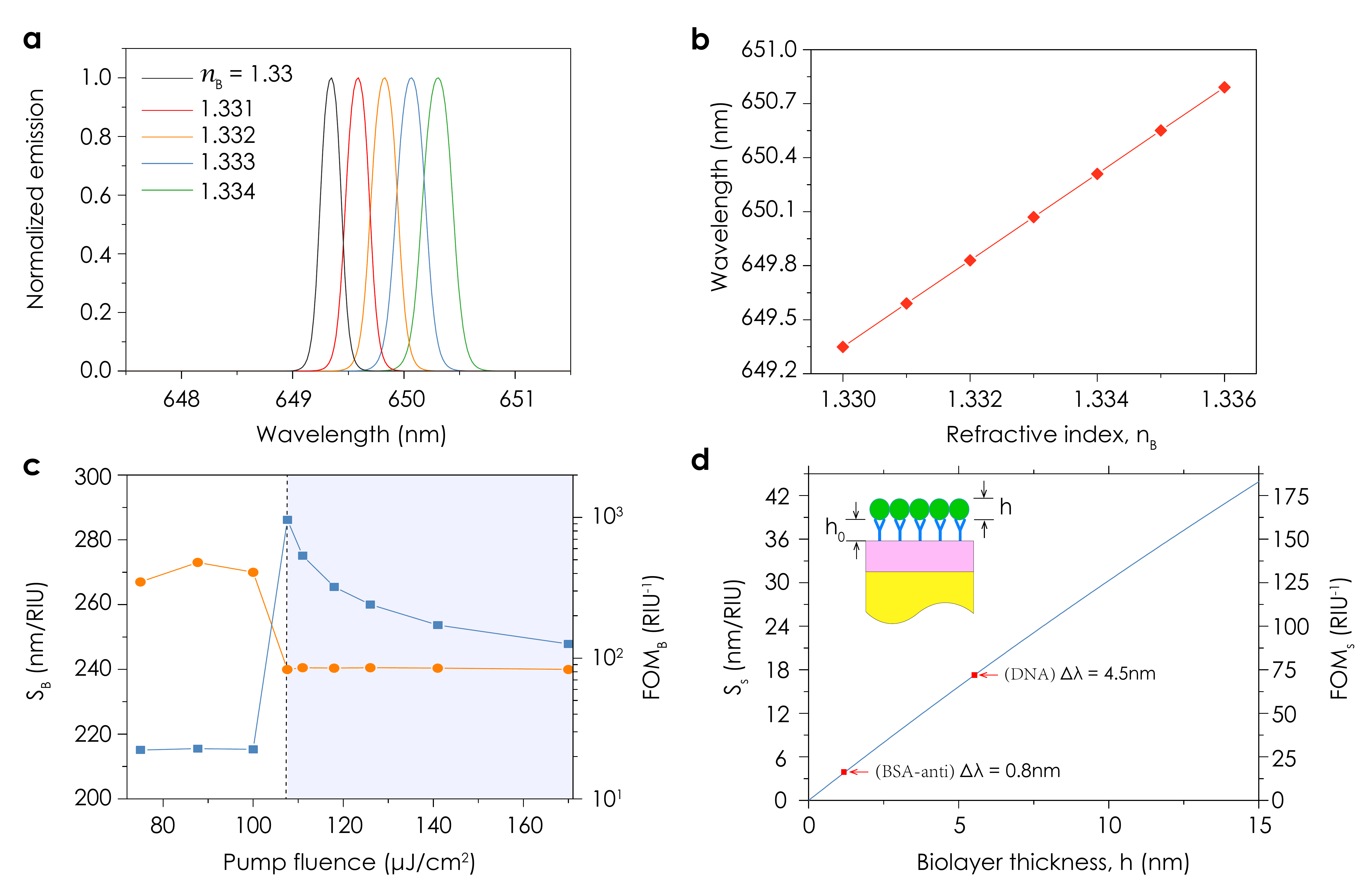}
\caption{Sensing performance of the active structure: (a) Spectral shift of laser emission as a function of the analyte's bulk
refractive index, $n_B$. (b) Wavelength of emission peak as a function of $n_B$. (c) Bulk sensitivity, SB, and bulk sensing figure of merit, FOM$_B$, as a function of the pump fluence. The vertical dashed line indicate the threshold condition. (d) Surface sensitivity, $S_S$, and surface sensing figure of merit, FOM$_S$, as a function of the biolayer thickness, $h$, for $h_0$ = 2.5 nm and $w = w_{th}$.}
\label{Sensing}
\end{figure*}

We focus first on the bulk sensing performance. For spectral interrogation, the bulk sensitivity is given by $S_B = \Delta \lambda = \Delta n_B$, where $\Delta\lambda$ is the spectral shift of the sensing signal (the peak lasing wavelength) and $n_B$ is the refractive index of the bulk analyte solution, which is close to the refractive index of water ($\sim$ 1.33). Thus, to obtain the value of S$_B$, we assume that the refractive index of the medium around the biosensing structure is in the range $1.33 < n_B < 1.336$. Fig.~\ref{Sensing}a and 4b show FDTD calculations of the emission spectrum and peak emission wavelength as a function of n$_B$ at the threshold pump fluence, $F_{th}$. We observe a linear shift in the peak emission wavelength and no detectable change in the spectrum's linewidth. Thus, the bulk sensitivity is S$_B$ = 240 nm/RIU, obtained as the slope of the curve in Figs.~\ref{Sensing}b, which yields a figure of merit of FOM$_B$ = S$_B$/$w_{th}$ = 1000 RIU$^{-1}$ at the lasing threshold condition.

The values of S$_B$ and FOM$_B$ as a function of the pump fluence are shown in Fig.~\ref{Sensing}c. Interestingly, we note that the S$_B$ is slightly larger below the lasing threshold, compared to the value above threshold. This is the result of a modest reduction in the electric field's decay constant at the lasing onset~\cite{Mazzotta2015}. Despite the this fact, the FOM$_B$ below threshold is quite small below threshold because of the large spectral linewidth. On the other hand, the FOM$_B$ peaks at the threshold condition and monotonically drops as the pump fluence increases because of the increasing spectral linewidth (see Fig.~\ref{Lasing}b). The FOM$_B$ value obtained at the lasing threshold is significantly larger than that of high performance passive plasmonic biosensors~\cite{Spackova2016} and experimentally demonstrated room temperature active plasmonic sensors using spectral interrogation~\cite{Wang2017a, Wang2017b}.

Now, we focus on the surface sensing performance. The surface sensitivity is described by $S_S = \Delta\lambda = \Delta n_S$, where $n_S$ is the refractive index change induced by an analyte biolayer of thickness $h$ located at a distance $h_0$ from the biosensor's surface. Furthermore, it is well established that this figure can be expressed as~\cite{Spackova2016}

\begin{equation}
S_S = S_B exp(-2h_0/\delta) [1-exp(-2h/\delta)] 
\end{equation}

\noindent where $\delta$ is the decay constant of the field into the analyte solution. The corresponding figure of merit is therefore FOM$_S$ = S$_S$/$w$. For definitiveness, we consider a value of $h_0$ = 2.5 nm corresponding to the monolayer thickness of a typical bioreceptor proteins, such as bovine serum albumin (BSA) or A/G protein~\cite{Wang2016}.

The values obtained for S$_S$ using Eq. (1) at the lasing threshold condition are shown in Fig.~\ref{Sensing}d. For this, we have used $w = w_{th}$ = 0.24 nm and $\delta$ = 142.8 nm, which corresponds to the decay constant averaged over the surface of the structure (see supplementary material). Note that despite the relatively large decay constant $\delta$, the value of S$_S$ can be as large as 40 nm/RIU within the first 15 nm above the bioreceptor layer. From these results, we can estimate the wavelength shift induced by the analyte biolayer as $\Delta \lambda$ = $S_S\Delta n_S$, with $\Delta n_S$ = 0.24 being the refractive index difference between water and a densely packed protein biolayer~\cite{Gao2013}. Thus, considering for instance a monolayer of BSA antibody (BSA-anti), whose thickness is h $\approx$ 1 nm, we obtain $\bigtriangleup \lambda \approx $0.8 nm. Similarly, a monolayer of a larger molecule such as DNA (h $\approx$ 6 nm~\cite{Piliarik2012}), would results in $\Delta\lambda \approx $4.5 nm. Clearly, the predicted spectral shifts are significantly larger than the spectral linewidth, indicating the possibility of sensing minuscule changes at the surface provided that the spectral properties of the laser remain stable.

The large values of $\Delta\lambda$ compared to w$_{th}$ reflect the large FOMS of the biosensor (right scale on Fig.~\ref{Sensing} d), which can reach values of 125 RIU$^{-1}$ for biolayers with a thickness in the order of 10 nm. This is contrary to the behavior observed in a passive plasmonic biosensor~\cite{Mazzotta2015, Svedendahl2009} because both, S$_S$ and $w$, are inversely proportional to $\delta$. Thus, while a small $\delta$ certainly results in a larger S$_S$ in a passive plasmonic biosensor, it does not necessary increases its FOM$_S$. On the other hand, the FOM$_S$ can be greatly increased in the active configuration because the narrow linewidth $w$ is dictated by the pump fluence (Figs.~\ref{Lasing}b and 4a) and does not influence the value of $\delta$. For comparison, we have estimated the FOM$_S$ values of typical passive plasmonic biosensors, assuming h = 10 nm and operation wavelength similar to that used in this work, obtaining the following values: 2.2 RIU$^{-1}$ for LSPR, 3.6 RIU$^{-1}$ for SPR with Kretschmann configuration~\cite{Spackova2019}, and 10 RIU$^{-1}$ for a structure based on lattice-plasmon resonance~\cite{Spackova2013}. Based on this observation we can anticipate that the active design studied here can offer an excellent surface sensing performance.

\section{Summary and conclusion}
In conclusion, we have investigated the sensing performance of an active biosensor based on a plasmonic crystal structure incorporating an organic gain medium. The plasmonic crystal supports a high-Q Bloch mode that favors plasmonic lasing with gain levels practically attainable by organic gain media at room temperature. The optical properties, and the bulk and surface sensing performance are investigated through extensive FDTD simulations incorporating the rate equations governing the atomic population dynamics of the gain medium. Our analysis indicates that lasing with a narrow linewidth of 0.24 nm is possible under practical optical pumping conditions. We showed that the narrow laser linewidth together with a large sensitivity result in FOM for bulk sensing as large as $\sim$ 1000 RIU$^{-1}$, which is at least an order of magnitude larger than that of high-performance passive plasmonic biosensors previously reported. We also show that FOM for surface sensing can be extremely large, reaching values of $\sim$ 65 RIU$^{-1}$ for thin biolayers in the order of 5 nm. We emphasize that a large surface sensing FOM is possible despite a relatively weak field confinement because, contrary to passive plasmonic sensors, the linewidth of the sensing signal is to a large extent independent of the plasmonic field confinement. The outstanding FOM values for bulk and surface sensing predicted here indicate that a high-$Q$ active sensing architectures, such as the studied here, have potential for low-concentration biomolecule detection. 

\section{Acknowledgment}
T.W. and J.S. are grateful to Prof. Renmin Ma for discussion. T.W. acknowledges the financial support from the National Natural Science Foundation of China (Grant No. 61804036), Zhejiang Province Commonweal Project (Grant No. LGJ20A040001), G.W. acknowledges National Key R\&D Program Grant (Grant No. 2018YFE0120000), and Zhejiang Provincial Key Research \& Development Project Grant (Grant No. 2019C04003). F.G. acknowledges National Natural Science Foundation of China (61801426), Zhejiang Provincial Natural Science Foundation (Z20F010018). I.D.L and Z.J. acknowledge the financial support from CONACyT Grant No. CN-17-109 and from the Federico Baur Endowed Chair in Nanotechnology.





\end{document}